 \documentclass[final,5p,times,twocolumn,authoryear]{elsarticle}

\usepackage{amssymb}
\usepackage{lipsum}

\journal{arxiv}

\begin{document}

\begin{frontmatter}



\title{The impact of the AI revolution on asset management}

\author[mkk]{Michael Kopp}
%\ead{michael.kopp@iarai.ac.at}
\address[mkk]{IARAI, Vienna}

\begin{abstract}
Recent progress in deep learning, a special form of machine learning, has led to remarkable capabilities machines can now be endowed with: they can read and understand free flowing text, reason and bargain with human counterparts, translate texts between languages, learn how to take decisions to maximize certain outcomes, etc. Today, machines have revolutionized the detection of cancer, the prediction of protein structures, the design of drugs, the control of nuclear fusion reactors etc. Although these capabilities are still in their infancy, it seems clear that their continued refinement and application will result in a technological impact on nearly all social and economic areas of human activity, the likes of which we have not seen before. In this article, I will share my view 
as to how AI will likely impact asset management in general and I will provide a 
mental framework that will equip readers with a simple criterion to assess whether and to what degree a given fund really exploits deep learning and whether a large disruption risk from deep learning exist.
\end{abstract}




\end{frontmatter}
\tableofcontents


\section{Introduction}
\label{introduction}
At its core, artificial intelligence (AI) promises to industrially scale `intelligence', including decision making, planning, understanding, reasoning with and about the world around us from digital datasets that world produces alone. 
On the back of the twin pillars of the `big data revolution' and ever increasing compute power (Moore's Law), the last decade has witnessed the emergence of stunning advances and real world applications of AI that are at the heart of a currently dominant belief that the underlying technology -- mostly `deep learning' -- will allow us to tackle hitherto unfathomable problems in nearly all areas of our lives.
Billions have been and are being invested into `deep learning' in 
sectors as diverse as drug discovery, food production, traffic management, 
map making, logistics, energy management, oil production etc. Such is the promise that it is only natural to muse how every industry is being and will be affected as this mushrooming technology unfolds and matures.
This article is my personal take on this question with regards to asset management. 
I have a somewhat unique view on the matter given my blended background. I am a former
portfolio manager responsible for managing capital in convertible bonds, merger arbitrage, credit, equity event driven and fundamentals based strategies. I am also a former mathematician who adopted deep learning in most aspects of his professional life early, building products and solutions with deep learning when it became capable enough, ran the research unit of a large mapping provider and am currently jointly running a research institute dedicated to AI. 
My motivation for writing this article is two-fold. Given my background, I face this question quite frequently from investors, fund managers but also technologists.
Moreover, I have the impression that the current discourse on the matter is somewhat
sketchy and could benefit from an opinion from my vantage point.
I will try and set out my view of what is in store for the asset management industry
in section 3 below. In section 2, I will give my take on where `deep learning' stands
today. In section 4, 
I will outline a useful framework, based on section 3, that allows one to classify
asset managers as to how they make use of deep learning today which also, in my view, highlights the disruption risk they will soon face from deep learning based approaches. 

\section{The deep learning revolution so far}

In this section I will argue that technological advances in artificial intelligence (AI) will have an undeniable
impact on all aspects of our life, that this technology is entirely different to other technology
advances and, hence, that applying it to real world problems successfully requires a novel approach of being aware of and embracing its idiosyncrasies.

\subsection{A technological revolution like no other}

Over the last decade, we have witnessed a remarkable technological revolution in which the dream of `artificial intelligence', as 
outlined in the famous 1956 symposium at Dartmouth college, has started to become a reality (see \cite{mccarthy2018proposal}).

\begin{figure}[t]
	\centering
	\includegraphics[width=8cm]{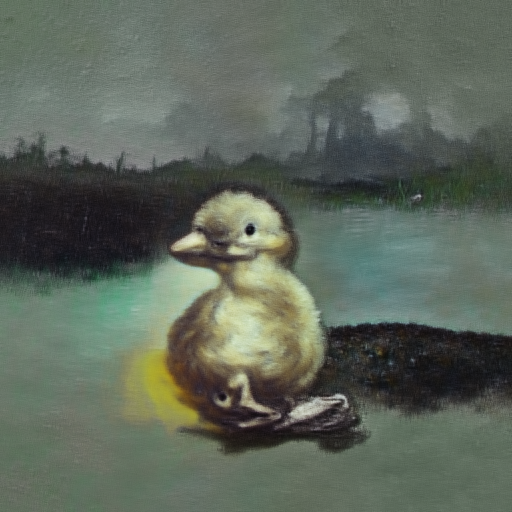}
	\caption{Example of image generated via a stable diffusion network from a CLOOB embedding from the prompt sentence 
	``a moody painting of a lonely duckling''}
	\label{fig1}
\end{figure}

Concretely, neural network based \citep{devlin2018bert,brown2020language, hoffmann2022training} large language models (LLMs) can now understand and generate text, in all languages, to such a 
degree that, recently, an engineer at Google
thought a new chat bot was scentient \citep{GoogleTuring2022} , thus clearly passing the famous Turing Test \citep{Turing1950machineIntelligence}.
Such models can even revive, decipher and explore hitherto extinct languages from few archaeological samples \citep{luo2019neural}. 
Deep learning has also had a profound impact on image analysis as well as generation. As examples of the former, facial recognition software is now scarily efficient, deep learning models beat 
pathologists at detecting cancer from cell samples \citep{cancer2021} and landslides can now be detected from simple satellite images from space \citep{ghorbanzadeh2022landslide4sense}.
Recent advances in new training regimes for neural networks have led to stunning advances in the latter. Where in the past large datasets needed to be annotated (labelled) 
for neural networks to teach themselves desired analytic skills by a large amount of examples, these networks can now learn relevant aspects of the data through either self-consistency
or having access to several different sensor readings on the same underlying observation. Surprisingly conceptually simple models like CLIP \citep{radford2021learning} and CLOOB \citep{furst2021cloob} can be trained on 
internet size datasets of pairs of images and text (basically images scraped from the web together with their caption), which lead to powerful joint feature embedding of text and image features
that allow for automatic image captioning as well as image generation from text as shown in Figure~\ref{fig1}.

Last, but not least, these same deep learning techniques have revolutionized control theory through `reinforcement learning'. Deep learning agents can now play complex computer games on
a super human level, including the game of go \citep{silver2016mastering}, DOTA2 against online players \citep{berner2019dota} and strategy and planning games like `No-Press Diplomacy'.
These advances are at the heart of novel electronic trading algorithms, the hope of achieving self-driving technology and emerging algorithms running nuclear fusion reactors as well as
electricity networks.

Combinations of all the above technologies has also recently lead to predicting the molecular structure of proteins \citep{AlphaFold2021} and are actively used in discovering novel
drugs \citep{CHEN20181241}. Moreover, as widely reported in the media, a merge of techniques from control theory and pre-trained generative large language models on chat room content has lead to chatGPT, a conversational AI agent which can answer questions, write computer programming code and also dream up answers that are false but plausible. 
% image recognition, annotation, generation.
% language (Google bot)
% agents that take action (trading, ...)
% Concrete references: 
% Google engineer scentient AI: https://www.washingtonpost.com/technology/2022/06/11/google-ai-lamda-blake-lemoine/
% LLM: BERT (devlin2018), GPT-3 (brown), Chincilla (hoffmann2022training),  
% CLIP, CLOOB (in bibtex: radford, fuerst)
% Stable Diffusion (rombach), GANs (goodfellow)
% supervised learning -> self-supervised learning, no labels needed.
% Foundation models (bommasani)
% techniques of utilizing such large trained foundation models and adapt them to real world tasks.
% Emergent abilities of large lanugage models (bibtex: wei)
% RL examples: (alphago: silver, 
% alphafold (alphafold), 
% Breast cancer detection (curroncol), drug discovery (CHEN), Landslide detection (ghorbanzadeh), traffic prediction (kopp)

\subsection{The odd nature of this deep learning revolution}

In order to understand how this technological revolution can be harvested, it is necessary to understand its 
peculiarities. Firstly, this revolution is founded on three pillars: data, compute power and heuristic exploration.
Modern deep learning models teach themselves from data alone (with or without labels). Almost all do so via simple gradient descent methods and the help of the chain rule (backpropagation method) which require the help of large, specialized
hardware accelerators (GPUs for instance). The key ingredient of designing these deep learning models is experimental trial and error that requires
both experience and mathematical guidance. This is so as there is no `theory' yet as to the architecture of a deep learning model 
that is supposed to work on a novel, hitherto not analyzed dataset. Rather, practitioners find models that worked on similar datasets 
and adjust them or compose several models that work on parts of the data that are known to capture relevant features.

There is a large part of research dedicated to methods as to how to combine/compose models (few-shot learning, meta learning,
student teacher architecture, etc.) and any advance in this area is immediately applicable in production.
For instance, the recently discovered ability to self train extremely large neural network models on certain massive, unlabelled data sources 
or a combination of such sources 
has led to the idea of having universal, reusable components, so called `foundation models' \citep{bommasani2021opportunities}
that are trained once and form the key lego-piece feature extractors for novel applications.
One unique aspect of this AI revolution the last sentence demonstrates is that fundamental research is turned into production code within weeks
or months, rather than years or decades as would be the case for most natural sciences.

The increased popularity AI is enjoying has lead to more trial and error in deep learning, which in turn, together with the increased 
experience of most AI researchers has led to an ever accelerating field. To put this into context, the last year has
seen such novel insights that most prior deep learning solutions are now obsolete again.

Last but not least, this latest leg in the deep learning revolution relies on internet size, unlabelled data being fed to models mostly trained to correctly guess masked out parts of the input. Given that a key component of the dominant neural network architecture (transformers) underlying most models these days as well as widely used generation techniques are known to be a powerful memory capable of storing and learning to retrieving large amounts of data \citep{ramsauer2020hopfield, carlini2023extracting}, it is no wonder that copyright lawsuits abound over whether training models on data off the internet really represents a `fair use' or infringement \citep{lawsuits2023}.
% data, compute and mathematically guided heuristic modelling.
% backward compatibility of pytorch: https://github.com/pytorch/pytorch/wiki/PyTorch's-Python-Frontend-Backward-and-Forward-Compatibility-Policy
% backward compatibility of tensorflow: https://www.tensorflow.org/guide/versions
% software 2.0: https://karpathy.medium.com/software-2-0-a64152b37c35
% NVIDIA dominant but secure - unclear what future hardware needs are. Currently backpropagation (i.e. `change rule' from calculus and gradient descent)
% cite (Hinton: ren)
% rapid technological progress.
% quick turn around from fundamental research to production.
% production step most critical and difficult.
% value shift.

\subsection{Consequences or why harnessing deep learning is not straight forward}

Building deep learning solutions in production is difficult given the `shape' of this AI revolution discussed above. 
\begin{itemize}
\item Which problems should be solved with deep learning and which ones should be tackled with traditional software? The former are black-box algorithms trained mostly by gradient descent. How are the boundaries handled?
\item Unlike most `traditional' technology solutions, deep learning models are hardly explainable and their performance can only really be gleaned from measurement (`black-box' nature).
\item Moreover, a shift in the underlying distribution of the data fed into a deep learning model compared to the data with which it was trained might lead to 
unexplained outcomes and potentially necessitate constant retraining and enhanced monitoring. 
\item Given that large scale trial and error is best achieved by community interaction and technology reuse, what should and should not be patented and what is the copyright situation of the training data used, i.e. the fuel on which these
models run?
\end{itemize}

A consistent set of answers to these questions is only just emerging and requires expertise in building these systems as well as understanding the problems one will face in the 
real world. This can be amply demonstrated by the fact that the key tool-chain on even writing code for AI models, the python frameworks tensorflow \citep{tensorflowbackward} 
and pytorch \citep{pytorchbackward}, 
are not backward compatible as each version will feature large changes necessitating changes in the programming languages' basic syntax. The rest of the usual tool chain (deployment, orchestration, etc.)
is even less fixed and also only just emerging.

\section{The shape of things to come in asset management}
%\lipsum[1-4]
The infamous quote `it is tough to make predictions, especially about the future'
is attributed to Yogi Berra and has been an uncomfortable truth befalling many predictions about the future. Nevertheless, even only considering the technologies
the current deep learning revolution has washed ashore and ignoring high level 
discussions and predictions about future successes, it seems hard to not come
to the conclusion that some vital parts of asset management can and will be heavily impacted, both in traditional human led proprietary trading as well as algorithmic trading. 
In what follows, I will thus ignore all well meant warnings, lean out of the window and attempt a deep glance into the crystal ball to glean the shape of things to come.

\subsection{Trade decision making in proprietary trading}
There are quite a few highly profitable and scalable trading strategies that have so far resisted being usurped by algorithmic trading. Usually, these depend heavily on a human level understanding of the underlying trade dynamic which is hard to explicitly capture with an algorithm. Take for example antitrust risk in a merger arbitrage deal. Assessing this risk requires an understanding of the competitive landscape such a merger would bring about, on the 
methodology and legal principles different antitrust authorities employ and the precise conditions the merger agreement sets out for a given set of such authorities for the deal to still go through. The emerging ability to build machines that can understand large corpora of texts and that can combine this understanding with past trading data of past deals in order to learn trading actions makes it rather plausible that a system can be built to assess this risk now. Such a system would need to reason about the acquirer's and target's business in different geographies, their competitors in these geographies usually marked out in regulatory filings, read thousands of merger documents, emerging legal opinions, verdicts and antitrust agency publications on methodologies and combine this information with trading data and other newspaper articles in the usual 
`rumourtage' outlets read by many merger arbitrage traders in order to form a quantitative view of the deal's antitrust risk, just like a human trader would do with her/his natural neural network. Yes, the task is not straight forward and all the caveats of the last section apply, but a technological stretch it seems to be not. In short, trading strategies that have hitherto withstood the onslaught of algorithmic trading are likely to come into scope.

\subsection{Automatically finding novel trading strategies}
Another fascinating capability that has emerged and is being exploited is the ability in control systems to find a plethora of different optimization strategies
satisfying complex rewards. Trading a portfolio with given imposed risk metrics is just such a scenario and hence it is now technologically feasible to create 
a large number of solutions with different characteristics for such a problem with the press of a button. In other words, with the press of a button a large number
of different artificial traders can be generated that all adhere to a given risk
limit whilst optimizing returns set over a given horizon.
We note that this ability to find new trading strategies in combination with other emerging deep learning capabilities is, in theory, enough to also disrupt existing algorithmic trading funds. Imagine a fund in which not only new short term or ultra short term trading strategies are found via deep learning, but where the same underlying algorithms also produce highly optimized code to actually execute
the trades, to connect to exchanges and products in a new way and to continuously improve all parts of the underlying code making up the entire trading operation.
This is, although technologically not infeasible today, unlikely to exist any time soon at this stage. That said, I am just taking existing pieces of deep learning successes and extrapolating their combination and so, total sci-fi it seems to be not to me.  

\subsection{Capital allocation}
Combining the ability of the last subsection with the ability of a learnt capital allocation system that maximizes rewards of a large number of automated trading strategies seems technologically feasible. In its simplest form, a deep learning system could learn, from reward data alone, to allocate capital at each step to traders whose past performance is known - reducing this to yet another control theory problem that seems solvable. A more sophisticated version would require being mindful of paradigm shifts not just of past data but a near to infinite number of dreamt up future scenarios is also within the realm of the possible, namely in model-based reinforcement learning and model-based imitation learning. Of course, such dreamt up scenarios have to be realistic enough and self-consistent enough to be helpful and current technologies
are only scratching the surface on this. Nevertheless, with enough compute power and data it would be hard to argue that this could not be achieved today in particular trading settings.

\subsection{Risk management}
This ability to dream up realistic future scenarios in large numbers and force deep learning agents to do well on these scenarios is also key in risk management. Such systems could read and comprehend every trade-able security's underlying risk as well as the relevant context of our world in an outright long only or long short portfolio, for instance, and could estimate the entire portfolio's reaction to certain shocks. Take for instance an unexpected closure of the Suez canal or the invasion of Ukraine. It is not infeasible with today's technology that large language models have stored enough data to be able to verbally reason that the Suez canal affects the shipping routes of certain raw materials and hence any security that is linked to such a material. That said, whether or not such dreamt up scenarios are in fact feasible and not just well constructed lies with current technologies is hard to say but mitigation approaches exist. Finally, although such advanced risk management on its own is clearly valuable, it seems equally clear that its value would be much larger if it was directly integrated at conception of any novel trading strategy as well as in the capital allocation process.   

\subsection{A word of warning}
The above predictions are, of course, idealistic and should be balanced by some realistic considerations. First and foremost, it is not at all clear that regulators or investors would cherish trading strategies where a human is taken completely out of the loop, even if that is feasible. More generally, the black box nature of current deep learning systems necessitate realistic back-testing or model-based approaches to assess the risk fully. That said, in a way this not different to the case of human traders who trade based on their natural neural networks and within strictly defined risk parameters that are constantly monitored.
Although such traders can verbalize their train of thoughts, it is hard to audit their risk any better than a deep learning trader envisaged here. Bayesian based algorithmic trading strategies, i.e. strategies which come with a theoretical return distribution rather than a single outcome that has been verified by back-testing for years are at a natural advantage here. That said, such strategies are usually of limited complexity and usually found in high frequency trading on ultra liquid assets where they are extremely profitable.

\subsection{A word about data}
Ominously absent in the above discussion is any word about the training data such systems would require. Apart from the emerging legal questions as to the copyright and fair use of data available on the internet as training data we commented on above, datasets ingested by deep learning systems for training purposes are usually
in the `dirty' raw format in which they occur in the wild especially if the datasets in question are large. This is somewhat in stark contrast to most other machine learning approaches where great care has to be taken to curate or label such datasets accurately at great cost. Moreover, the true value of automatic trading strategy discovery lies in the fact that such strategies automatically decide what data is relevant. The only constraint for throwing arbitrary amounts of data at a model is compute power, but practical techniques exists to efficiently
encode data neurally (through either fixed or learnt representations). The emergence of efficient scaling compute and dataset provision techniques \citep{pytorchbackward, svogor2022profiling} facilitate that task.

\section{A framework for assessing the role of deep learning in a given fund}
%\lipsum[1-4]
It is clear on reading the last subsection and ordering its thoughts that by far the biggest impact of deep learning on asset management will come from the ability to automate finding trading strategies and capital allocation decisions and, necessarily, a form of risk assessment (either learned implicitly or broken out separately or, best, both for redundancy). This would not only allow for such deep learning driven funds to start competing in strategies where, currently, human traders (aided by quantitative or AI driven tools) excel, but also come up with new
trading strategy paradigms that have so far not emerged. Successes in AI over the past years have shown that both such `super-human' successes are feasible and such super-human strategies are in fact intuitive to us if they hail from deep learning 
algorithms that were explicitly inspired by human like learning. A stunning example of the latter is the fact that the only game the human world champion Lee Sedol won against AlphaGo in the game of Go in 2016 was him replaying a trick on the machine he learned from it in his first defeat to it. This means he was able to abstract a principle behind a sequence of moves he observed and apply that principle on a new setting, fooling the originator of that principle in this new setting. This seeming weakness of the machine is actually a strength, for further training of AlphaGo with today's more powerful compute resources would have reduced that risk but also, transferable strategies are more explainable strategies -- an important aspect for regulatory reasons.

\subsection{A simple criterion}
The above insight leads to a rather simple classification criteria for funds across asset and trading strategy classes that captures where deep learning can have an impact: the level of automation of trading and of capital allocation. This results in the following 4 broad classes one can bucket any fund manager into.
\begin{itemize}
    \item[] {\noindent \bf Level 0:} trading decisions made by human traders (likely aided by quantitative or AI driven tools such as Bloomberg terminals) and capital allocation which follows a static rule or is dynamic and result dependent.
    \item[] {\noindent \bf Level 1:} automated trading decisions with trading strategies designed and validated with a human in the loop and either a static
    or dynamic capital allocation rule also designed with a human in the loop.
    \item[] {\noindent \bf Level 2:} automated trading decisions and automated discovery of trading strategies conceived and designed without a human in the loop and a human designed or in the loop capital allocation rule as well as trade supervision.
    \item[] {\noindent \bf Level 3:} fully automated and human independent capital allocation and trading strategy discovery and, necessarily risk management.
\end{itemize}

\subsection{Remarks}
Technologically, Level 3 asset managers are now within the realm of the possible and I know of at least one fund that is tantalizingly close to achieving this goal. 
One challenge of course is regulatory compliance: after all, Level 3 trading strategies employed could be anything and operators of such a fund manager would have to figure out the underlying strategy via introspection, as well as the learned risk management associated underlying capital allocation. A challenge operators of such a fund would have would be to assure that the underlying algorithm has also learned its disclosure obligations towards potential regulators which can be both be sophisticated and vary widely between jurisdictions. For instance finding and combining several public data sources that predict large market moving trades in the equity market of others and exploiting the resulting market moves might be `front running' in some jurisdictions but not in others.
It is therefore likely, and for the one instance of a fund I know in fact the case, that such Level 3 funds will first emerge trading largely unregulated assets (such as crypto currencies) extremely successfully.
Most quantitative hedge fund managers would fall into Level 1 or Level 2. I have my doubts a Level 3 strategy can be achieved without a deep learning approach such as the one outlined above.
I note in passing that Level 1,2 and 3 fund managers do not necessarily need to execute trades automatically, rather the decision to trade as well as the strategy underlying these decisions are fully automated.
The purpose of this simple categorization scheme really is to show where deep learning based methods, mostly or partially in existence today, can have a disruptive impact. In my view asset managers with a lower category level are at a higher risk of being disrupted by deep learning based approaches. Of course, one could run the argument that, for instance, successful traditionally human traders who invest in, say, equity long only value strategies could also use deep learning techniques to discover new strategies and, should they be intuitive enough to be recognizable, filter them and employ them just as Lee Sedol did in his winning game against AlphaGo. My issue with this argument is that with increased computing power, the level of `super-human-ness' of the trading strategies is, in my view, likely to increase and our human ability to ascribe a strategy might reach its limits, too. Maybe one could get the machine to literally spell out its reasoning, but at some point even that might not help bridge this gap if the compute power employed is large enough.

\section{Summary and conclusions}
%\lipsum[1-4]
In section 2 I argued that deep learning is currently the dominant driving technique of the AI revolution and that its odd nature requires adopters of these
incredible techniques to embrace their oddness. In section 3 I have taken a deep look into the crystal ball and sketched how I think already existing techniques can and will end up impacting the asset management industry. A resulting observation is then explored in section 4, namely that deep learning can have a tremendous impact on trade strategy discovery automation and automatic capital allocation and using this as a simple classification criterion for asset managers captures, in my view, their true adoption of deep learning so far as well as the disruptive risk they face from deep learning techniques. 
One point I wish to stress about the above is the necessity to embrace the odd nature of deep learning I highlight is fundamental and not easy to achieve in asset management. Investors and regulators need to be comfortable with the black-box nature of this technology which also implies that it should be viewed more human like rather than a defined algorithm in terms of its ability to err. Deep learning is incredibly successful not least because the ultimate underlying philosophy giving rise to it was for these algorithms to be human like. Little wonder then that they make mistakes like we do. Deep learning and AI based traders are like human ones and require the organizational setup we place around human traders in order to mitigate risk, so, say, a deep learning based risk-manager that cuts positions and is trained in an adversarial way, etc. To my mind, the challenge to mount a truly deep learning based asset manager requires both a culture to embrace its black-box-nature in all aspects as well as the ability to be a tail blazer with investors and regulators for this technology. This is a large effort, often underestimated in my view, but likely extremely profitable for the first ones to achieve this. 

\section*{Acknowledgements}
Thanks to my colleagues at the institute of advanced research in artificial intelligence (IARAI), specifically Sepp Hochreiter, David Kreil, Moritz Neun and Alina Mihai as well as all my academic collaborators in AI. Without their help, hard work and discussions I could not have formed the opinion expressed above.
I would also like to thank my former colleagues Kaveh Sheibani, Julian Harvey Wood and Philippe Lamarque and everyone at Pendragon Capital in its various iterations for teaching me all I know 
about traditional proprietary trading and portfolio management. Last, but not least, I would like to thank Urs Alder for incredibly helpful suggestions on earlier drafts of this paper. It would not be in existence without his kind and expert help. 
%Lastly, but not least, I would like to thank R+S Capital Management, specifically Kirill Rubinstein
%and Thomas Schmiedel for blazing a trail in fund management with AI and for taking me
%along for the ride.

\bibliographystyle{elsarticle-harv} 
\bibliography{main}




%%\bibitem[ ()]{}

%%\end{thebibliography}
\end{document}